\documentclass[twocolumn]{revtex4-1}

\usepackage{braket}
\usepackage{graphicx}

\usepackage{amsmath}

\makeatletter
\let\@fnsymbol\@fnsymbol@latex
\@booleanfalse\altaffilletter@sw
\makeatother

\begin{document}

%\title{Full title: Experimental realization of the quantum walks of a radial local phonon in a linear crystal of trapped ions\\ 
%Short title: Quantum walks of a phonon in trapped ions}
\title{Quantum walks of a phonon in trapped ions}

% \author{Masaya Tamura}
% \affiliation{Graduate School of Engineering Science, Osaka University.}
% \author{Takashi Mukaiyama}
% \affiliation{Graduate School of Engineering Science, Osaka University.}
% \affiliation{%
% Quantum Information and Quantum Biology Division,
% Institute for Open and Transdisciplinary Research Initiatives,
% Osaka University.}
% %\author{Kenji Toyoda}%
% %\email[]{Corresponding author: toyoda@qiqb.otri.osaka-u.ac.jp}
% \author{Kenji Toyoda%
% \thanks{Corresponding author: toyoda@qiqb.otri.osaka-u.ac.jp}}
% \affiliation{%
% Quantum Information and Quantum Biology Division,
% Institute for Open and Transdisciplinary Research Initiatives,
% Osaka University.}

\author
{%
Masaya Tamura,$^{1}$ Takashi Mukaiyama,$^{1,2}$ Kenji Toyoda$^{2\ast}$}
\affiliation{
%\\\rm
$^{1}$Graduate School of Engineering Science, Osaka University, %\newline%
Toyonaka 560-8531, Japan.\\%\hspace{1cm}\\%\newline
%}
%\affiliation{
$^{2}$Quantum Information and Quantum Biology Division,
Institute for Open and Transdisciplinary Research Initiatives,
Osaka University, %\newline
Toyonaka 560-8531, Japan.\\%\hspace{1cm}%\newline
%}
%\affiliation{
%\\
\rm $^\ast$To whom correspondence should be addressed; E-mail:  toyoda@qiqb.otri.osaka-u.ac.jp.}

\date{\today}

\begin{abstract}

Propagation and interference of quantum-mechanical particles comprise an
important part of elementary processes in quantum physics, and their
essence can be modeled using a quantum walk, a mathematical concept that
describes the motion of a quantum-mechanical particle among discretized
spatial regions.  Here we report the observation of the quantum walks of
a phonon, a vibrational quantum, in a trapped-ion crystal.  By employing
the capability of preparing and observing a localized wave packet of a
phonon, the propagation of a single radial local phonon in a four-ion
linear crystal is observed with single-site resolution.  The results
show an agreement with numerical calculations, indicating the
predictability and reproducibility of the phonon system.  These
characteristics may contribute advantageously in advanced experimental
studies of quantum walks with large numbers of nodes, as well as
realization of boson sampling and quantum simulation using phonons as
computational resources.

\end{abstract}


\maketitle
\parskip=0pt

\section{Introduction}

Propagation of quantum-mechanical particles in media
is essential to various fields including
quantum optics, condensed-matter physics and quantum information science.
Recent experimental progress has enabled observing
the propagation of individual quanta, and this is discussed
in the context of quantum walks (QWs) %
\cite{Aharonov1993,Venegas-Andraca2012}.
QWs can be extended to multiple walkers \cite{Peruzzo2010},
and are used to model universal computation \cite{Childs2009,Childs2013}.
So far photons have been a major platform for realizing QWs %
\cite{Bouwmeester2000,Perets2008,Peruzzo2010}. 
There have been also realizations with
neutral atoms \cite{Preiss2015} and trapped ions 
\cite{Schmitz2009,Zahringer2010}.
More recently, QWs of one and two correlated microwave photons
have been demonstrated with a superconducting system \cite{Yan2019}.

Local phonons (quanta of local vibrational motions) 
in the trapped ions \cite{Porras2004} show particle-like characteristics
and undergo interference
via beamsplitter-like coupling due to inter-ion Coulomb interactions 
\cite{Porras2004,Brown2011,Harlander2011,Haze2012,Toyoda2015},
thereby acting in a similar way to photons.
Phonons in trapped ions has the merit
of being generated and detected in deterministic ways.
Optical manipulation of phonons and incorporating interaction among
them are possible \cite{Debnath2018,Ivanov2009,Toyoda2013,Ding2017a,Ding2017b}.
These characteristics can be utilized in realization of QWs as
well as boson sampling \cite{Shen2014} and the Jaynes-Cummings-Hubbard model 
\cite{Ivanov2009,Toyoda2013}.

Propagation of local phonons so far has been observed with
axial motions of two ions in double-well potentials or
radial motions of a two-ion linear crystal  
\cite{Brown2011,Harlander2011,Haze2012}, and in a three-ion linear crystal
where blockade of hopping using site-dependent AC Stark shifts
is realized \cite{Debnath2018}.
In \cite{Ramm2014,Abdelrahman2017}, 
energy transport is studied with radial motional
modes in chains of up to 42 trapped ions which are cooled to
thermal states using Doppler cooling.

In this article, we present results on the propagation of a local phonon
among four sites of an ion crystal.
The radial motion of the ion crystal is cooled to near the ground state,
and one phonon is prepared at a desired site with an individual 
optical access,
whose propagation over the ion crystal is traced by a site-resolved 
observation. 
The phonon freely propagate among the four ion sites,
showing the high-contrast patterns of complex wave-packet interference
for up to 10 ms, which corresponds to $\sim$60 times 
the maximum adjacent-hopping time
(the maximum time required for a phonon to move between adjacent two sites).
The interference patterns are compared with numerically calculated results,
showing agreements in detail.
The experimental results are also analyzed in the frequency domain,
showing agreements with theoretical predictions
based on radial collective-mode analyses.

The phonon propagation observed here, 
which is interpreted as a continuous-time QW \cite{Farhi1998}
of a local-phonon wave packet,
can be extended in a straightforward manner
to larger numbers of ion sites, and hence the investigation of 
continuous-time QWs in larger systems may become possible.
Extension with regard to the number of phonons is also
possible, allowing the realization of boson sampling for 
the demonstration of quantum computing power that may overwhelm
the state-of-the-art classical computing
\cite{Aaronson2011,Shen2014}.

\section{Results}

\subsection{QWs of a phonon}

\begin{center} 
\begin{figure}[tbp]
\includegraphics[clip,width=14cm]{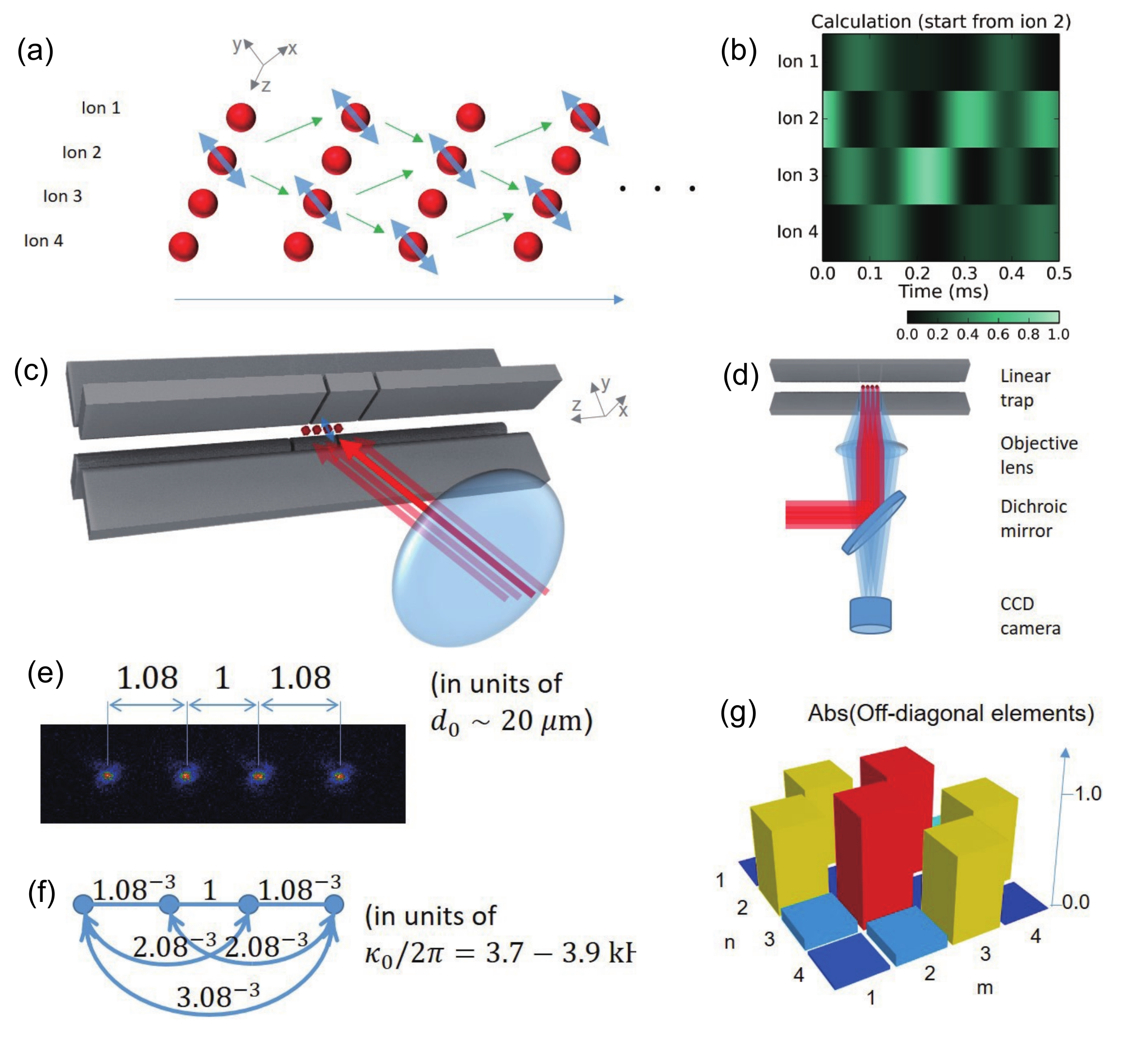}%
\caption{
\label{fig:Concepts}
{\bf QWs of a phonon.}
(a) Conceptual diagram for the QW of a phonon. 
First a phonon (radial local phonon oscillating in the $y$ direction) is 
prepared in the ion 2. 
Due to the Coulomb coupling, it split into two wave packets and
propagates to either the ion 1 or 3.
The ion 1 is at the edge of the crystal, and hence the
the phonon wave packet is reflected back there,
propagating again to the ion 2, where it interferes with
another wave packet coming from the ion 3.
In this way the complex pattern of propagation and interference is
formed.
(b) Numerically calculated result for the QW of a phonon.
A phonon is prepared in the ion 2, from which it propagates and interfere
with itself as described above.
Conditions similar to what is used in the experiment are assumed 
in this calculation.
(c) Setup for the optical excitation of the ions to generate and 
observe a phonon.
A linear Paul trap used, four ions trapped inside it, optical beams for 
excitation and an objective lens are depicted. 
(d) Schematic for the setup of optical excitation and fluorescence 
observation.
(e) Ion string used. The distances between adjacent pairs of ions
are described.
(f) Graph structure when the phonon propagation is
viewed as a continuous-time QW.
The values are hopping rates corresponding to each edges
in units of $\kappa_0/2\pi$ (hopping rates between the central two ions). 
(g) Absolute values of the Hamiltonian off-diagonal elements 
for phonon propagation. 
}
\end{figure}
\end{center}

Figure \ref{fig:Concepts} shows the concepts, 
configurations and conditions for the experiment on 
the QWs of a phonon.
Figure \ref{fig:Concepts}(a) shows 
the conceptual diagram for the QWs of a phonon
in a linear crystal of four ions (numbered as 1-4).
First a phonon (radial local phonon oscillating in the $y$ direction) is 
prepared in the ion 2. 
Due to the inter-mode coupling arising from the Coulomb interaction
between ions, the phonon split into two wave packets and
propagates to either the ion 1 or 3.
Since the ion 1 is at the edge of the crystal, 
the phonon wave packet that has proceeded there gets reflected and
propagates back to the ion 2, where it interferes with
another wave packet coming from the ion 3.
In this way a complex pattern of propagation and interference is
formed.
A numerically calculated result showing a similar behavior is given in 
Fig.\ \ref{fig:Concepts}(b).
Conditions similar to what is used in the experiment are assumed 
in this calculation.
The ions are illuminated with four independent beams 
[Fig.\ \ref{fig:Concepts}(c)],
and fluorescence from all ions is imaged onto an electron multiplying 
charge-coupled device [Fig.\ \ref{fig:Concepts}(d)].
% Fig.\ \ref{fig:Concepts}(c) 
% shows the setup for the optical excitation of the ions to generate and 
% observe a phonon.
% In a linear Paul trap four ions are trapped, and optical beams for 
% excitation are applied through an objective lens.
% Fig.\ \ref{fig:Concepts}(d) shows the
% schematic of the setup for optical excitation and fluorescence 
% observation.
Figure \ref{fig:Concepts}(e) shows the
ion string used, where the distances between adjacent pairs of ions
are described.
Figure \ref{fig:Concepts}(f) shows the
graph structure when the phonon propagation is
viewed as a continuous-time QW.
The values are the hopping rates corresponding to each edges
in units of $\kappa_0$ (hopping rates between central two ions). 
The corresponding Hamiltonian off-diagonal elements 
(see Materials and Methods) are plotted in Fig.\ \ref{fig:Concepts}(g).
% Figure \ref{fig:Concepts}(g) shows the
% absolute values of the Hamiltonian off-diagonal 
% elements for phonon propagation
% (see Materials and Methods). 

\subsection{Results in the time domain}

\begin{figure}[h]
\begin{center}
\includegraphics[clip,width=18cm]{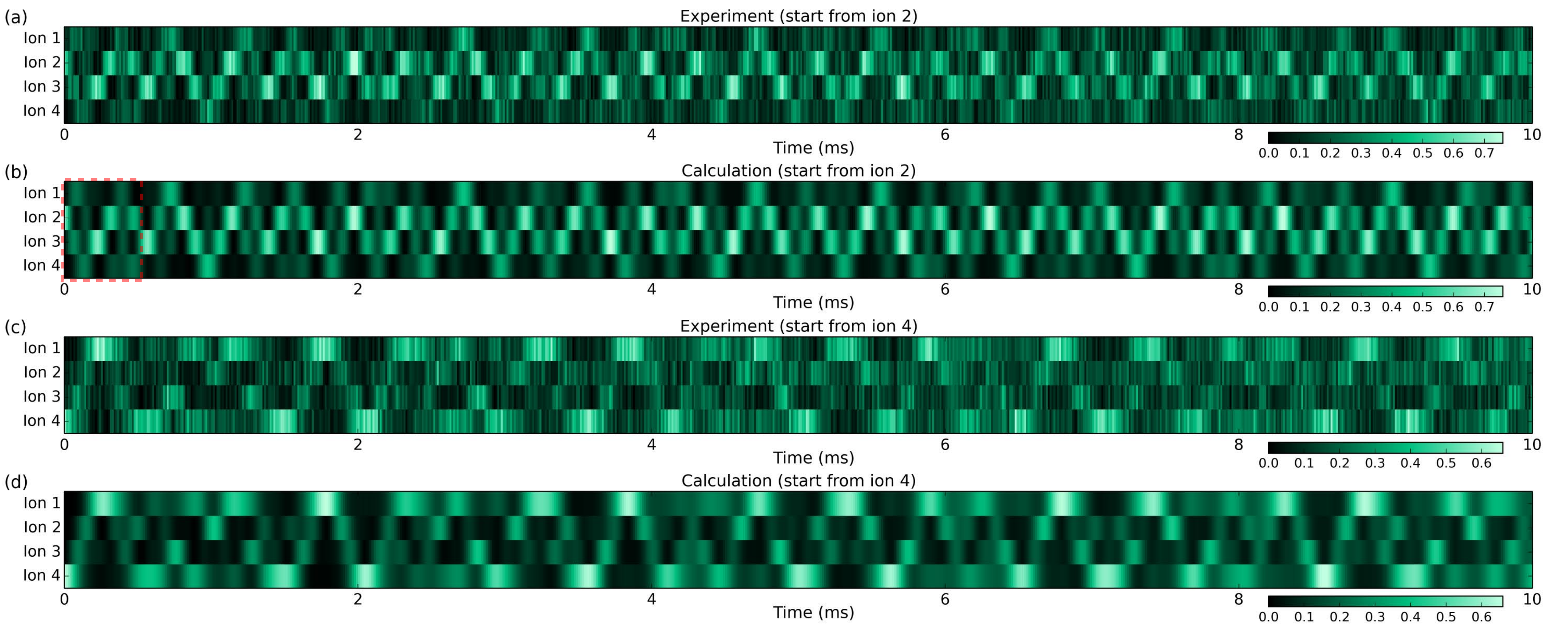}%
\caption{
\label{fig:time-domain}
{\bf Results for phonon propagation in the time domain.}
The experimental and numerically calculated results for the time dependence
of the local-phonon existence probability
after one phonon is prepared at a particular site are displayed as brightness.
(a) [(c)] Experimental result with the initial phonon population at the ion 2 [4].
Each of the four rows represent the time dependence of
the local-phonon existence probability
at each ion site, from the ion 1 (top) to 4 (bottom).
The width of each time step is 12.5 $\mu$s, and 
the number of measurements per step is 50.
(b) [(d)] Numerically calculated result with the initial phonon population at the ion 2 [4].
The area surrounded by the thin-red dashed 
rectangular at the left end in (b)
corresponds to that displayed in Fig.\ \ref{fig:Concepts}(b).
}
\end{center}
\end{figure}

Figure \ref{fig:time-domain}(a) (\ref{fig:time-domain}(c)) shows the experimental result
for the phonon propagation
with the initial phonon population at the ion 2 (4).
The phonon prepared at the ion 2 (4) splits
and propagates among the four ion sites.
Figure \ref{fig:time-domain}(b) (\ref{fig:time-domain}(d)) shows the numerical result 
with the initial phonon population at the ion 2 (4)
calculated
using the Hamiltonian of Eq.\ \ref{eqn:Hamiltonian-actual}
in Materials and Methods.
For obtaining these calculated results, the following four parameters
are adjusted so as to minimize the residual sum of squares:
the horizontal time scale 
%(determined by $\kappa_0$, 
%which is determined by radial and axial confinements after all), 
%determined by radial and axial confinements after all), 
(determined by $\kappa_0\propto\omega_z^2/\omega_y$ where
$\omega_z$ and $\omega_y$ are the oscillation frequencies for
the axial $z$ and radial $y$ directions, respectively),
the horizontal offset (due to the fact that hopping may occur during
the preparation time), the population scale factors (explained below), 
and the heating rate ($\sim$5 quanta per second).
The population scale factors are the overall proportionality factors
multiplied to the calculated results.
We need to adjust those factors because of the reduction
of preparation and observation efficiency due to the decoherence in
applying the red-sideband $\pi$ pulses.
Their explicit values 
are 0.66 and 0.76 for the initial phonon population at the ion 2 and 4,
respectively.
Each calculated result shows an agreement with the corresponding
experimental result up to the propagation time of 10 ms.

\begin{figure}[h]
\begin{center}
\includegraphics[width=8.5cm]{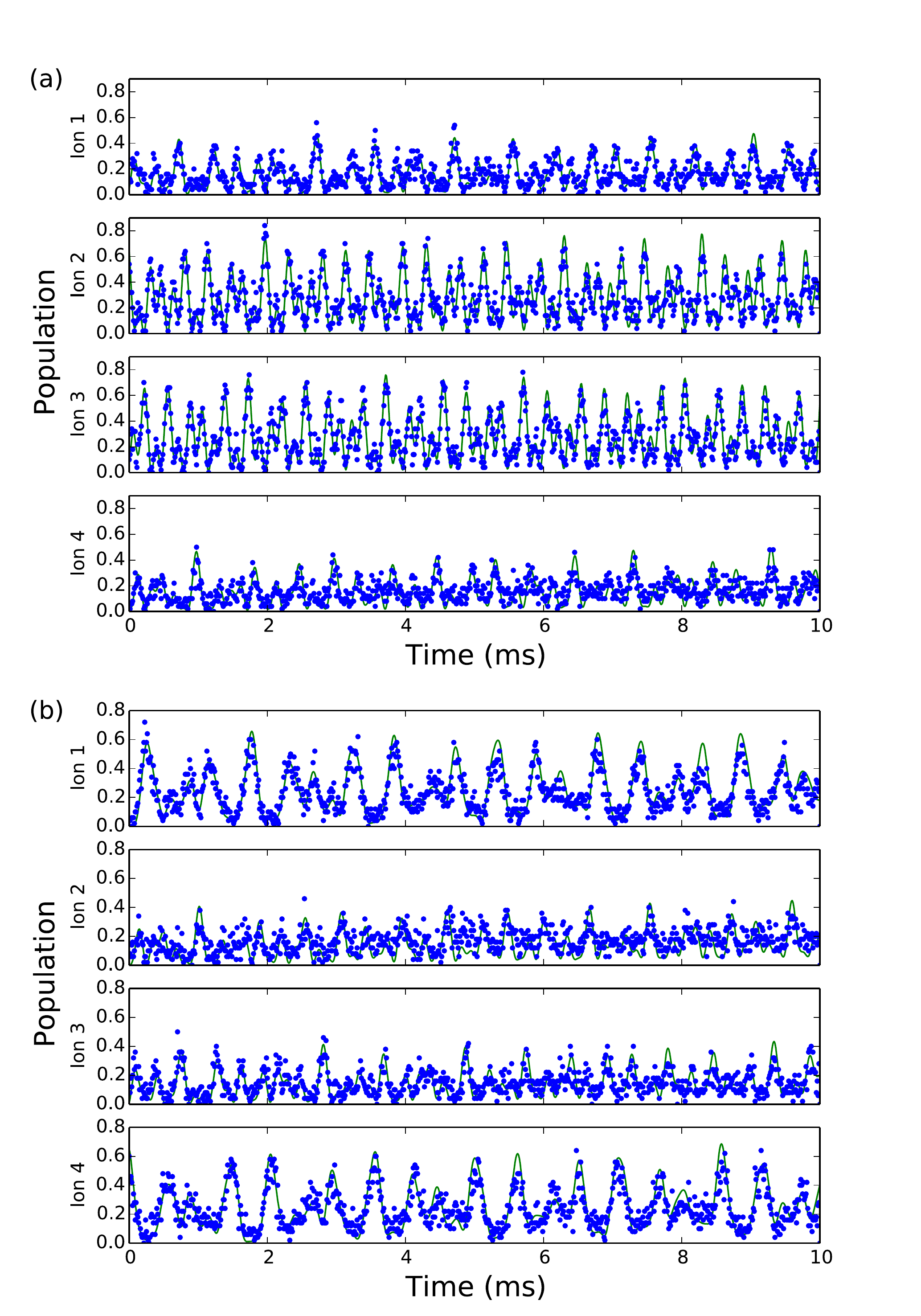}\\%
\caption{
\label{fig:time-domain-2D}
{\bf Two-dimensional plots of the results for phonon propagation in the time domain.}
The same experimental and numerically calculated results 
as shown in Fig.\ \ref{fig:time-domain}) are displayed
using two-dimensional plots.
(a) [(b)] Result with the initial phonon population at the ion 2 [4].
Each of the four rows represent the time dependence of
the local-phonon existence probability
at each ion site, from the ion 1 (top) to 4 (bottom).
The points represent the experimental data points, and
the curves represent the numerically calculated results.
For the experimental data points,
the width of each time step is 12.5 $\mu$s, and 
the number of measurements per step is 50.
% In the numerically calculated results,
% heating rates of $\sim$5 quanta per second, which give the best matching
% with experimental results (the minimum residual sum of squares), 
% are assumed.
}
\end{center}
\end{figure}

Figure \ref{fig:time-domain-2D} shows the
experimental and numerically calculated results in the time domain
(the same results as shown in Fig.\ \ref{fig:time-domain})
displayed using two-dimensional plots.
Figure \ref{fig:time-domain-2D}(a) (\ref{fig:time-domain-2D}(b)) shows the 
result with the initial phonon population at the ion 2 (4).
Each of the four rows represent the time dependence of
the local-phonon existence probability
at each ion site, from the ion 1 (top) to 4 (bottom).
The points represent the experimental data points, and
the curves represent the numerically calculated results
(obtained by adjusting the parameters to minimize 
the residual sum of squares, as described above).

\subsection{Results in the frequency domain}

Since the local phonon can be interpreted as a superposition of 
collective-mode phonons,
the behavior of phonon propagation in the time domain can be
understood as a wave interference among different collective modes. 
From this view point, it is expected that
the Fourier transforms of the time-domain results
give information about the collective modes involved,
thereby enabling the analyses of the phonon propagation in the frequency domain.
Below we describe the formalism useful for such analyses.

By diagonalizing the potential and Coulomb-energy parts of the 
radial-motion Hamiltonian with the harmonic approximation, 
eigenvalues $\omega_p$ and real-valued eigenvectors $b_n^{(p)}$
can be obtained \cite{Zhu2006}, where
$p$ and $n$ are the indices for the 
radial collective modes and ions, respectively.
%can be obtained,
%where $n$ is the index for ions \cite{Zhu2006}.
The Heisenberg-picture local-phonon annihilation operator at the ion $n$
can be represented as follows using $b_n^{(p)}$ and the collective-mode
annihilation operator $\hat{c}_p$:
\(\hat{a}_n(t)=\sum_p{}b_n^{(p)}\hat{c}_p{}e^{-i\omega_p{}t}\).
The initial state with one phonon at the ion $n$ is
\(\ket{\Phi_n}=\hat{a}_n^\dagger(0)\ket{\Phi_0}
=(\sum_p{}b_n^{(p)}\hat{c}_p^\dagger)\ket{\Phi_0}\),
where $\ket{\Phi_0}=\ket{0,0,\cdots,0}$ is
the ground state of the collective-mode Fock states.
Using these, 
the probability for one phonon (prepared at time 0 and the ion $n$)
observed at time $t$ and ion $m$ 
is obtained as follows:
\begin{align}
&P_{nm}\equiv\braket{\Phi_n|\hat{a}_m^\dagger(t){}\hat{a}_m(t)|\Phi_n}
\nonumber\\
&=\sum_{p,q}b_n^{(q)}b_m^{(q)}e^{i\omega_qt}e^{-i\omega_pt}b_m^{(p)}b_n^{(p)}
\nonumber\\
&=\sum_{p}\left(b_n^{(p)}b_m^{(p)}\right)^2
+\sum_{q>p}b_n^{(q)}b_m^{(q)}b_m^{(p)}b_n^{(p)}
\left[e^{i(\omega_q-\omega_p)t}+\rm{c.c.}\right]\nonumber
\end{align}

Therefore, the amplitude of the 
positive and negative frequency component oscillating 
at $\omega_q-\omega_p$ and $-(\omega_q-\omega_p)$, respectively, 
is equal to $b_n^{(q)}b_m^{(q)}b_m^{(p)}b_n^{(p)}$
($q>p$ and $\omega_q>\omega_p$ for $q>p$ assumed).

\begin{figure}[h]
\begin{center}
\includegraphics[width=8.5cm]{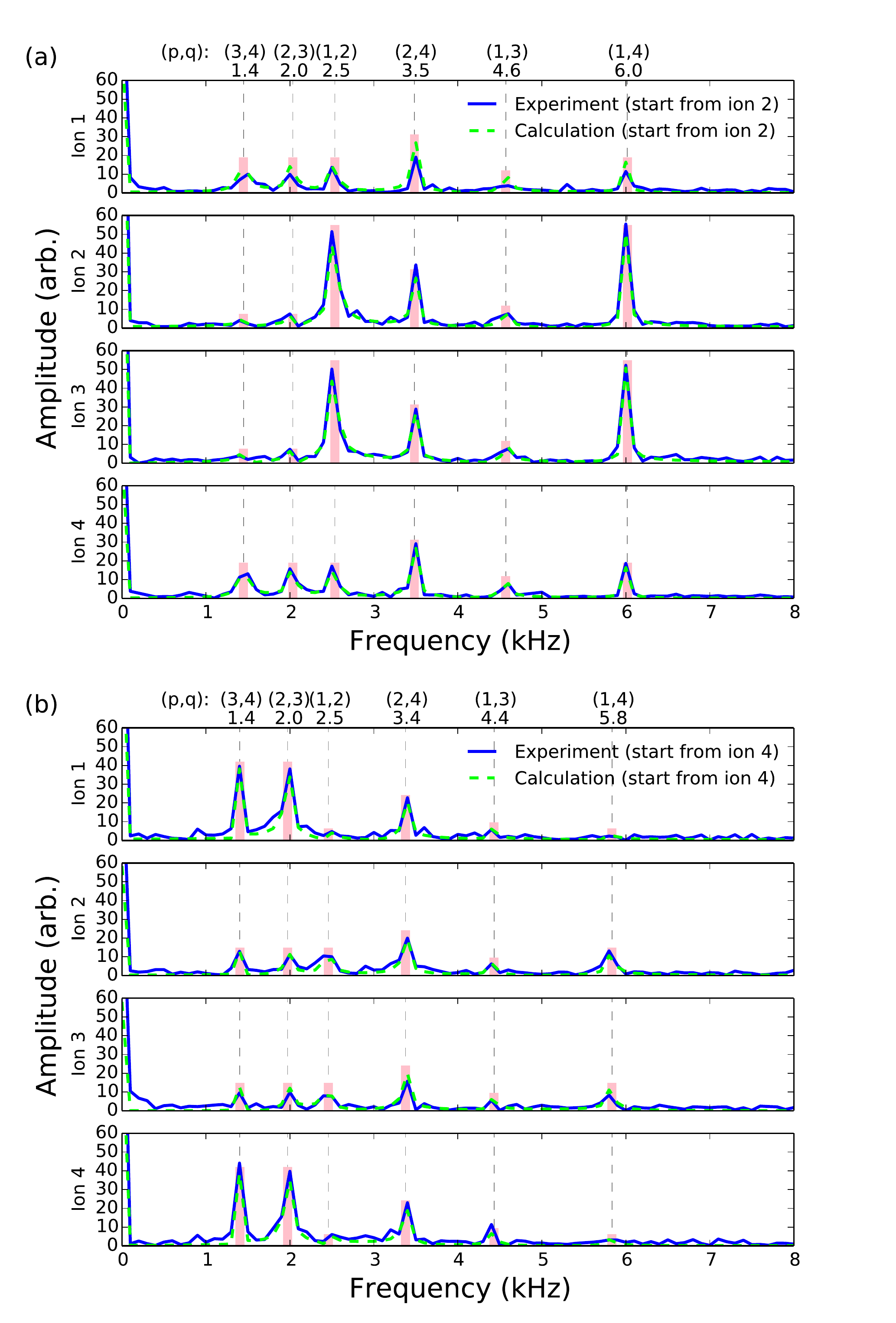}\\%
\caption{
\label{fig:freq-domain}
{\bf Results for phonon propagation in the frequency domain.}
(a) [(b)] Result with the initial phonon population at the ion 2 [4].
The Fourier transform of the local-phonon existence probability
is plotted.
The blue solid (green dashed) curves represent 
the experimental (numerically calculated) results.
The light-red vertical bars represent analytically obtained results 
corresponding to $b_n^{(q)}b_m^{(q)}b_m^{(p)}b_n^{(p)}$ (see text).
The vertical gray dashed lines represents the difference of 
frequencies ($\omega_q-\omega_p$) for each pairs of radial collective modes. 
The texts at the top represents the pairs of indices $(p,q)$
for the relevant collective-mode pairs, where the indices 1-4
are taken in the order of increasing oscillation frequencies.
The texts in the second row from the top represents
the difference of frequencies ($\omega_q-\omega_p$)
in the unit of kHz.
}
\end{center}
\end{figure}

In Fig.\ \ref{fig:freq-domain},
the Fourier transforms of populations at the ion sites
are shown.
The solid blue curves are the discrete Fourier transforms of the 
experimental results, and the
green dashed curves are those of the calculated results.
The light-red vertical bars are the analytically obtained results
$b_n^{(q)}b_m^{(q)}b_m^{(p)}b_n^{(p)}$
multiplied with appropriate proportionality factors
for comparison with the discrete Fourier transforms.
The vertical gray dashed lines represents the difference of 
frequencies ($\omega_q-\omega_p$) for each pairs of radial collective modes. 
It is understood from the plot that the three results match well 
with each other.

\section{Discussion}

%In summary, we have 

In conclusion, 
we have experimentally shown that a phonon in trapped ions
undergoes propagation and interference that can be interpreted as a quantum walk. We have observed propagation dynamics for up to 10 ms,
and the results has been compared with numerically calculated ones,
showing agreements.
It can be expected that the system of phonons offers a platform 
for realizing quantum walks in larger scales and 
non-universal quantum-computationa schemes including 
boson sampling. 

The current experiment is performed with single phonon,
while it has been pointed out that
single particle (discrete) QWs are shown to be implementable
with classical devices \cite{Knight2003a,Knight2003b,Jeong2004,Venegas-Andraca2012}.
The genuine quantum mechanical properties can be investigated with 
the use of multiple walkers.
Extension to multiple phonons is also possible in the system of phonons
in trapped ions.
Preparation of multiple phonons in desired sites
in a deterministic way is possible with standard techniques
using sequences involving carrier and sideband transitions
and site-resolved optical manipulation.
The detection of multiple phonons (projection measurements) 
is also possible in principle,
and has been performed with a single ion \cite{An2015}.

The efficiency of the initial-state preparation and final-state measurement
in our system is limited due to the infidelity in red-sideband $\pi$ pulses
($\sim$0.7-0.8 when combined).
Since the present experiment is performed using 
a relatively weak axial confinement and tight focuses, 
the residual thermal fluctuation along the axial 
direction ($\sim$0.5 $\mu$m rms) 
is nonnegligible compared with the size of the focuses ($\sim$3 $\mu$m).
Other possible factors of the decoherence might be 
the combinations of beam jitter,
deteriorated spatial modes at the focus 
and AC Stark shifts \cite{Haffner2003}.

Although heating is observed in the time span of up to 10 ms, 
the decoherence of phonon dynamics is not prominent
in this time scale.
Previously decay time of 13 ms, corresponding to $\sim$30 round trips
in the case of two ions has been observed \cite{Toyoda2015}, 
and this decay time
seems to be dependent on certain experimental conditions,
which include the radial/axial confinements, 
and the residual thermal motion along the axial direction,
which is cooled only with Doppler cooling.

\section{Materials and methods}

\subsection{Phonon propagation as QWs}
Propagation of a local phonon can be viewed as a QW.
A QW is defined as a temporally homogeneous
evolution of a quantum system defined on a graph
comprising of vertices and edges.
Two different formulations of QWs exist, namely 
in discrete time \cite{Aharonov1993} and continuous time \cite{Farhi1998}.
It can be shown that local phonon propagation in trapped ions
is equivalent to a continuous-time QW.

Continuous-time QWs are formulated using
a generator matrix $M_{nm}$ ($\gamma$ is the jump rate
and $k_n$ is the number of edges that are connected to vertex $n$):
\(M_{nm}=
-\gamma_{nm}	\ (n\neq{}m, n \mbox{ and } m \mbox{ connected by an edge});
0			\ (n\neq{}m, n \mbox{ and } m \mbox{ not connected});
\sum_{l\neq{}n}\gamma_{nl}	\ (n=m)\).
The Hamiltonian $\hat{H}$ with matrix elements given
by $\braket{n|\hat{H}|m}=M_{nm}$ ($\hbar{}=1$)
generates an evolution described with
the Schr\"odinger equation and the unitary evolution operator
$U=\exp(-i\hat{H}t)$,
which is interpreted as a QW in continuous time.

The Hamiltonian for a radial direction (denoted as the $y$ direction) 
in a linear ion crystal
is described as follows \cite{Porras2004}:
\begin{equation}
 \hat{H}_y=
\sum^N_{n=1}(\omega_y+\omega_{y,n})\hat{a}_i^\dagger\hat{a}_n-
\sum^N_{m>n}t_{nm}%
(\hat{a}_n^\dagger\hat{a}_m+\hat{a}_n\hat{a}_m^\dagger).
\end{equation}
Here
\(t_{nm}=-e^2/8\pi\epsilon_0m\omega_y|z_n^0-z_m^0|^3\)
is the hopping amplitudes and 
\(\omega_{y,n}=\sum^N_{m\neq{}n}t_{nm}\)
is the harmonic correction terms for radial oscillation frequencies.
If the rotating frame oscillating at $\omega_y$ is adopted
and the case of only one phonon is considered
($\hat{a}_n^\dagger\hat{a}_m\rightarrow\ket{n}\bra{m}$),
\(\hat{H}_y=\sum^N_n(\sum_{m\neq{n}}t_{nm})\ket{n}\bra{n}-
  \sum_{m\neq{n}}t_{nm}\ket{n}\bra{m}\).
By setting $\gamma_{nm}\equiv-t_{nm}$, this corresponds exactly to
continuous-time QWs explained above.

\subsection{Experimental conditions}

The experiment is performed with four $^{40}$Ca$^+$ ions 
(named as the ion 1-4) trapped in 
a three-dimensional linear Paul trap.
The oscillation frequencies for the three directions $(x,y,z)$ are
$(\omega_x,\omega_y,\omega_z)/2\pi\sim(3.1,2.9,0.09)$ MHz, respectively.
The distance of the central two ions ($d_0$)
is $\sim20$ $\mu$m. 
One of the radial directions referred to as $y$
is mainly used for the observation of phonon propagation.
The radial-phonon hopping rate between the central two ions
defined as
$\kappa_0/2\pi=(e^2/4\pi\epsilon_0m\omega_yd_0^3)/2\pi$ is 
3.7 - 3.9 kHz.
The heating rate for $y$ estimated from the propagation results is 
$\sim$5 quanta per second.
Due to this heating, 
%in order to assure that up to only a single quantum of phonons
%exist in the system,
we limit the total measurement times up to 10 ms
to make sure that only a single quantum of phonons
exist in the system.
The excess phonon population due to heating can be suppressed
to below 5 \% in this case.

In the condition given above, 
the explicit values for the Hamiltonian matrix elements
are obtained as follows.
\begin{equation}
 \frac{\kappa_0}{2}
\left(
  \begin{array}{rrrr}
   -0.93&0.79&0.11&0.03\\
   0.79&-1.90&1.00&0.11\\
   0.11&1.00&-1.90&0.79\\
   0.03&0.11&0.79&-0.93\\
  \end{array}
 \right)
\label{eqn:Hamiltonian-actual}
\end{equation}
The smallest matrix elements among the 
adjacent couplings in this Hamiltonian is $0.79\times\kappa_0/2$.
This determines the maximum adjacent-hopping time,
that is, the maximum time required for a phonon to hop
to an adjacent site, which is obtained to be
$(0.79\times\kappa_0/2\pi)^{-1}/2\sim160$ $\mu$s.
By using this, we can conclude that
the total observation time used in this work (10 ms)
corresponds to $\sim$60 times the maximum adjacent-hopping time.

The $S_{1/2}(m_J=-1/2)$--$D_{5/2}(m_J=-1/2)$ transition at 729 nm 
is used as the
main transition for optical manipulation 
and observation of phonons.
Each of four beams at 729 nm is focused onto each ion
with the $1/e^2$ radius of $\sim$3 $\mu$m.
The power for each beam is adjusted so that
the Rabi frequency for every ion become equal to each other,
which amounts to 
$\sim$400 kHz for the carrier transition and $\sim$16 kHz
for the blue-sideband transition.
This equal illumination is important for assuring equal efficiencies
of preparation and mapping (explained in the next subsection) for all ions.
Each of the four beams can be switched by using a dedicated 
acousto-optic modulator.

\subsection{Time sequence}
First the ion crystal is Doppler cooled in all directions,
and the radial two directions ($x$ and $y$) 
are cooled with sideband cooling to 
$\{\bar{n}_x,\bar{n}_y\}\sim\{0.1,0.03\}$.

Then a blue-sideband and carrier $\pi$ pulses 
($\sim$31 $\mu$s and $\sim$1.3 $\mu$s, respectively) 
are applied to a selected ion,
which produce one radial local phonon.
This is followed by a free evolution period with variable length, 
in which the generated phonon hops among the ion sites.
After this, a mapping pulse (red-sideband $\pi$ pulse, $\sim$31 $\mu$s) 
is applied,
which is followed by an internal state measurement using a 397-nm laser
(resonant to the $S_{1/2}$--$P_{1/2}$ transition), whereby 
the two internal states $S_{1/2}$ and $D_{5/2}$ are discriminated.

\begin{acknowledgments}
%This work was supported by MEXT Q-LEAP.
{\bf General:} 
We thank Yutaka Shikano for valuable discussions.
{\bf Funding:} 
This work was supported by MEXT Quantum Leap Flagship Program (MEXT Q-LEAP) 
Grant Number JPMXS0118067477.
{\bf Author contributions:}
M.T. conducted the experiment.
T.M. and K.T supervised the project.
M.T. and K.T. analyzed the data and performed numerical calculations.
K.T. prepared the initial version of the manuscript.
All authors discussed the results and contributed to
the writing of the manuscript.
{\bf Competing interests:}
The authors declare that they have no competing interests.
{\bf Data and materials availability:}
All data needed to evaluate the conclusions in the paper are present in
the paper and/or the Supplementary Materials. Additional data related to
this paper may be requested from the authors.
\end{acknowledgments}

\bibliography{Phonon_propagation}

\onecolumngrid

\end{document}